\documentclass[11pt,a4paper]{article}
\usepackage{jheppub}
\usepackage{amsmath}
\usepackage[most]{tcolorbox}
\usepackage{dsfont}
\usepackage{ulem}
\usepackage{natbib}
\usepackage{xcolor}
\usepackage[hang,flushmargin]{footmisc}
\usepackage{tikz-cd}
\usepackage{enumitem}
\usepackage{amsfonts,amssymb, amscd,amsmath,latexsym,amsbsy,bm}
\usepackage{stmaryrd}
\usepackage{todonotes}
\usepackage{stackrel}
\usepackage{amsthm}

\usepackage{graphicx}

\title{Witten'ın kübik açık sicim alan teorisinde analitik çözümler üzerine}

\author{İlmar Gahramanov$^{K,B,c}$}

\affiliation{
$^K$Fizik Bölümü, Boğaziçi Üniversitesi, 34342 Bebek, İstanbul, Türkiye\\[-0.4cm]

$^B$Radyasyon Problemleri Enstitüsü, B.Vahabzade 9, AZ1143, Bakü, Azerbaycan

$^c$Matematik Bölümü, Khazar Üniversitesi, Mehseti St. 41, AZ1096, Bakü, Azerbaycan
}

\abstract{Bu makale, 2024 yılında ``Feza Gürsey Fizik Günleri'' okulunda verdiğim derse dayanmaktadır. Makalede, takyon vakumu için açık kübik sicim alan teorisinin analitik çözümleri (özellikle Erler-Schnabl çözümü) kısaca gözden geçirilmektedir.}

\dedicated{Durmuş Ali Demir'in\\
anısına}

\begin{document}

\maketitle

\section{Giriş}

\textit{2023 yılının Aralık ayında, Yıldız Teknik Üniversitesi Fizik Kulübü tarafından düzenlenen “Fizik Konferansı”nda, Durmuş Demir hocayla birlikte konuşmacıydık. Öğle arasında Durmuş Hoca, gitmesi gerektiğini ancak üzerinde çalıştığı yeni bir proje için \cite{Demir:2023hkm} benimle açık sicim alan teorisi üzerine konuşmak istediğini söyledi. Bu alanda analitik çözümler ilgisini çekmiş gibi görünüyordu. Kısa bir süre içinde tekrar buluşmak üzere anlaştık.}

\textit{Ancak 24 Şubat 2024 günü, Türk bilim dünyasını derinden sarsan acı haberi aldım: Durmuş Demir vefat etmişti. Maalesef, onunla konuşma fırsatımız olmadı. Uzun süredir açık sicim alan teorisi üzerinde çalışmıyordum, ancak Durmuş Hoca’yla konuşmayı planladığımız konuları öğrencilerle paylaşmak adına, “Feza Gürsey Fizik Günleri” kapsamında Durmuş Ali Demir’in anısına bir konuşma yaptım. Bu makale, o konuşmanın yazıya dökülmüş halidir.}\\

\vspace{0.2cm}

1986 yılında Edward Witten Chern-Simons tipi eyleme sahip açık bozonik sicim alanı teorisini önerdi \cite{Witten:1985cc}. Sicim alanı teorisi, sicim teorisinin pertürbatif olmayan bir tanımıdır. Sicim alan teorisi için çeşitli yaklaşımlar vardır \cite{Sen:2024nfd} ve bunlardan en başarılılarından biri Witten'in bu teorisi ve onun süpersimmetrik versiyonudur. Witten'in teorisinin eylemi aşağıdaki gibi yazılmaktadır
\begin{equation}
S \ = \ \frac12 \int \Psi \ast Q_B \Psi +\frac{g}{3} \int \Psi \ast \Psi \ast \Psi \;.
\end{equation}
Burada $\Psi$ alanı tüm fiziksel uzay-zaman alanlarını içeren sicim alanıdır
\begin{equation}
\Psi \ =\  \left[ \phi(x) + A_{\mu} \alpha_{-1}^{\mu} + \psi(x) c_0+\ldots \right] c_1 |0> \;.
\end{equation}
Eylemdeki yıldız $\ast$ çarpımını makalede tanımlamayacağız; detaylarla ilgilenenler bu makalelere \cite{Bonora:2009zz,Bonora:2013txj} bakabilir.


Bu eylemin minimizasyonundan elde edilen hareket denklemi
\begin{equation}
    Q_B \Psi + \Psi \ast \Psi =0
\end{equation}
kolay görünmesine rağmen çözümlerinin bulunması zor bir problemdir ancak son 20 yılda  Witten'in açık sicim alanı teorisine analitik çözümler bulma konusunda önemli ilerlemeler kaydedildi. Analitik çözümlere ilişkin gelişmelerin kapsamlı incelemeleri bu makalelerde bulunabilir \cite{Fuchs:2008cc, Schnabl:2010tb, Okawa:2012ica, Erler:2019vhl, Erbin:2021smf, Sen:2024nfd}.

2005 yılında takyon vakuumunu karakterize eden ilk analitik çözüm Schnabl tarafından bulundu. Takyon yoğunlaşmasına yönelik analitik bir çözüm, Sen'in varsayımlarını \cite{Sen:1999xm} kontrol etmek için güçlü bir araçtır\footnote{Aslında Schanbl'ın çözümü birinci ve ikinci varsayımlarını doğrulamaktadır.}. Sen'in ilk varsayımına göre, takiyon yoğunlaşması $D25$-zar'ın yok olmasına karşılık gelmelidir, yani kararlı ve kararsız arka plan arasındaki enerji yoğunluğu tam olarak $D25$-zar gerilimini vermelidir.
\begin{equation} 
T_{25} = (2 \pi^2 g^2)^{-1}
\end{equation}
Daha sonra Erler ve Schnabl yeni bir çözüm teklifinde bulundular ve analitik çözümler dönemi başladı. Takyon yoğunlaşmasına yönelik analitik çözümleri incelemenin bir diğer motivasyonu da çözümlerin kozmolojik modellerdeki uygulamalarıdır. Bu yöndeki bazı uygulamalar \cite{Forini:2005bs, Calcagni:2005xc, Arefeva:2010zza} makalelerinde bulunabilir.

Bu yazıda Erler ve Schnabl \cite{Erler:2009uj} tarafından 2009 yılında bulunan klasik çözümü ve onun özelliklerini kısaca gözden geçireceğiz ve bu çözüme benzer bir çözümden konuşacağız.

\section{Analitik çözümler}

Erler-Schnabl çözümünü yazmak için öncelikle $\{K, B, c\}$ alt cebirini tanıtmamız gerekmektedir. Bunun için aşağıdaki üç alanı $K$ (Grassmann çift), $B$ (Grassmann tek) ve $c$ (Grassmann tek) tanımlayalım
\begin{equation} \label{1}
K= \frac {\pi}2 \left(\frac12 K_1+\frac{1}{\pi} ({\cal L}_0+{\cal L}^\dagger_0)\right)|I\rangle,\quad\quad B=\frac {\pi}2 \left(\frac12 B_1+\frac{1}{\pi} ({\cal B}_0+{\cal B}^\dagger_0)\right)|I\rangle, \quad\quad c= c(1) |I\rangle  \;,
\end{equation}
burada $K_1, {\cal L}_0$ ve $B_1, {\cal B}_0$ dünya yüzeyinin (worldsheet'in) enerji-momentum tensörü ve $b$ hayalet (ghost) alanı cinsinden aşağıdaki şekilde ifade edilir
\begin{align}
{\cal L}_0 & = L_0+\sum_{k=1}^{\infty} \frac{2 (-1)^{k+1}}{4k^2-1}L_{2k} \; ,\\
{\cal B}_0 & = b_0+\sum_{k=1}^{\infty} \frac{2 (-1)^{k+1}}{4k^2-1}b_{2k} \; ,\\
K_1 & = L_1 + L_{-1} \; ,\\
B_1 & = b_1 + b_{-1} \;.
\end{align}
Böylece $K,B,c$ sicim alanları, aşağıdaki komütasyon ilişkileri ve BRST varyasyonları ile bir cebir oluşturur
\begin{align}
& [K,B]=0,\quad\quad [B,c]_+ =1, \quad\quad [K,c]= \partial c,\quad\quad [B,\partial c]_+=0 \\ 
& Q\,B=K,\quad\quad [Q,K]=0,\quad\quad, Q\, c=cKc\label{KBc}
\end{align}

$KBc$ cebiri sicim alan teorisinde analitik çözümleri oluşturmak ve incelemek için çok yararlı bir araçtır, bkz. \cite{Erler:2009uj, Erler:2012qn, Murata:2011ex}.

Artık Erler--Schnabl çözümünü yazmaya hazırız. Hayali terim içermeyen bu klasik açık sicim alan teorisi çözümü aşağıdaki şekildedir \cite{Erler:2009uj}
\begin{equation}
\psi_{ES} = c B(K+1)c \frac 1{K+1}
\end{equation}

Bu çözüm, pertürbatif vakumun tekil gauge dönüşümü yoluyla bu şekilde de  yazılabilir\footnote{Ayrıntılar için bkz. \cite{Okawa:2006vm, Erler:2012qn}.}
\begin{equation} 
\psi_{ES} \ = \ U_0 Q U_0^{-1} \;,
\end{equation}
burada
\begin{equation}
U_0=1-\frac{1}{K+1}Bc \;\; \text{ve} \;\; U_0^{-1} = 1+\frac{1}{K}Bc \;.
\end{equation} 

Girişte bahsettiğimiz gibi takyon vakum çözümü kullanılarak Sen'in varsayımları kontrol edilebilir. Erler--Schnabl çözümünün Sen'in ilk varsayımını karşıladığını göstermek için bu çözüme karşılık gelen enerjiyi hesaplayalım. Bu çözüm eksi $D25$--brane gerilimi\footnote{$g=1$ alıyoruz.} yeniden üretmektedir. Gerçekten enerji yoğunluğunu hesaplarsak
\begin{align} \nonumber
E[\psi_{ES}]&=\frac 16 \langle \psi_0, Q\psi_0\rangle= \frac 16 \langle (c+cKBc) \frac 1{K+1}\,cKc\,\frac 1{K+1}\rangle \\ 
&= \frac 16 \langle c \,\frac 1{K+1}\,cKc\,\frac 1{K+1}\rangle - \langle Q\left(Bc \frac 1{K+1}\,cKc\,\frac 1{K+1}\right)\rangle \;.
\end{align}
Son terim, BRST tam olduğu için yok olur ve ikinci satırdaki ilk terim şu ifadeyi verir:
\begin{align}  \nonumber
E[\psi_{ES}] &= \frac{1}{6} \int_0^\infty dt_1 dt_2 \, e^{-t_1-t_2} \langle c\, e^{-t_1 K}\,c \partial c \,e^{-t_2K}\rangle_{C_{t_1+t_2}} \\ \nonumber
&= -\frac{1}{6} \int_0^\infty dt_1 dt_2 \, e^{-t_1-t_2} \, \frac{(t_1+t_2)^2}{\pi^2} \sin^2\left(\frac{\pi t_1}{t_1+t_2}\right) \\ \label{Epsi01}
&= -\frac{1}{2\pi^2},
\end{align}
burada  $C_t$, $\arctan$ çerçevesindeki çevresi $t$ olan bir silindiri ifade eder. Aynı zamanda yukarıdaki ifadelerde Schwinger parametrizasyonu kullanmaktayız 
\begin{equation}
\frac{1}{1+K} = \int_{0}^{\infty} e^{-t(1+K)} \;.
\end{equation}
(\ref{Epsi01}) ifadesinde birinci satırdan ikinciye geçerken, üst yarı düzlemde iki nokta korelatöründe
\begin{equation}
\langle c(z_1) c\partial c(z_2)\rangle = -(z_1-z_2)^2.
\end{equation}
$\xi = \arctan(z)$ dönüşümüyle bu ifade, $C_\pi$ silindirinde bu ifade elde ediliyor
\begin{equation}
\langle c(\xi_1) c\partial c(\xi_2)\rangle = -\sin^2(\xi_1-\xi_2).
\end{equation}
Son olarak, bu ifadeyi $\xi \to \frac{\ell}{\pi} \xi$ ölçeklendirmesiyle $C_\ell$ silindirine dönüştürmek gerekir.

Erler–Schnabl çözümü açık sicim durumlarını desteklemez, yani bu çözüm Sen'in ikinci hipotezi ile uyumludur. Belirli bir $\psi_0$ çözümünün açık sicim durumlarını desteklemediğini göstermek için aşağıdaki koşulu sağlayan $A$ homotopi operatörünü  bulmak gerekir\footnote{Ayrıntılı bir açıklama için bkz. \cite{Ellwood:2006ba}.}
\begin{equation}
Q_{\psi_0} A = 1,
\end{equation}
burada
\begin{equation}
Q_{\psi_0} A = Q A + \psi_0 A + A \psi_0.
\end{equation}
Gerçekten, Erler–Schnabl çözümü için aşağıdaki şekilde bir homotopi operatörü bulunmaktadır \cite{Ellwood:2006ba}
\begin{equation}
A = B \frac{1}{1+K}.
\end{equation}

Erler–Schnabl çözümü, Okawa tarafından önerilen bu formda yazılabilir \cite{Okawa:2006vm}
\begin{equation}
\psi = F(K) c \frac{KB}{1-F(K)^2}cF(K),
\end{equation}
burada $F(K)$, sadece $K$ sicim alanının bir fonksiyonudur. Bu form, bir formal çözüm sınıfını temsil eder ve $F(K) = \frac{1}{\sqrt{1+K}}$ durumu Erler–Schnabl çözümünü verir. Son ifade, aynı zamanda aşağıdaki şekilde bir gauge dönüşümüyle yazılabilir
\begin{equation}
\Psi = U^{-1} Q U \;\; \text{ve}\;\; U = 1 - F(K)cBF(K).
\end{equation}

2014 senesinde Loriana Bonora ile birlikte \cite{BonoraGahramanov} Erler--Schnabl çözümüne benzer ilginç bir analitik çözüm elde ettik. Yeni bir çözüm aramak için aşağıdaki ansatz'ı kullanalım
\begin{equation} 
\psi= (c(K+1)Bc) \frac {f(K)}{K+1} \;. 
\end{equation} 
Böylece, \( Q\psi = cKc \frac {f(K)}{K+1} \) ve \( \psi\psi = c\left(f(K) c - c f(K)\right) Bc \frac {f(K)}{K+1} \) elde ederiz. 
\( f(K) = a + bK \) olarak belirlersek, hareket denkleminin \( a - b = 1 \) şartını sağladığını görürüz. Bu durumda  
\begin{equation} 
\psi= \psi_b= (c(K+1)Bc) \left(\frac 1{K+1}+b\right) \;, \label{psib} 
\end{equation} 
şeklinde olur ve burada \( b \) herhangi bir sayı olabilir. (\ref{psib}), yeni bir klasik çözüm sınıfını ifade eder.  

Bu çözümlerin takyon vakum çözümleri olup olmadığına bakalım. Bu çözümlere karşılık gelen enerji yoğunluğunu hesaplayarak Sen'in birinci varsayımını kontrol edelim. Enerji hesaplaması, ek bir terim haricinde daha önceki hesaplama ile aynıdır:  
\begin{equation} 
b\,\langle c \frac 1{K+1}\,cKc\rangle =b\int_0^\infty dt \, e^{-t} \frac {t^2}{\pi^2} \sin^2\left(\frac {\pi t}t\right)=0 
\end{equation} 
Bu nedenle, tüm çözümler aynı enerjiye sahiptir ve takyon vakumunu temsil ederler.  

Daha önce söylediğimiz gibi analitik bir çözüm tekil bir gauge dönüşümü olarak yazılabilir. Gerçekten de,  
\begin{equation} 
U= 1-cB \frac {h}{K+1}, \quad\quad U^{-1}=1+cB \frac h{K+1-h}, \label{gauge} 
\end{equation}  
şeklinde tanımlasak (burada $h=h(K)$, sadece $K$ sicim alanın bir fonksiyonudur)
\begin{equation} 
U^{-1}Q U= c \frac {K+1}{K+1-h}BKc \frac h{K+1} 
\end{equation} 
ifadesini elde ederiz. Bu formülde \( h(K) = 1 + \beta K \) olarak alırsak  
\begin{equation} 
U^{-1}Q U= c (K+1)Bc\, \frac h{1-\beta} \frac 1{K+1}=  c (K+1)Bc\frac  1{K+1} \left(\frac 1{1-\beta}+ \frac {\beta}{1-\beta}K\right) 
\end{equation} 
ifadesi (\ref{psib}) çözümünü verir (burada $a= \frac 1{1-\beta}$ ile $b=\frac {\beta}{1-\beta}$ olarak tanımlamak gerekir). Bu genel çözümde $b=\beta=0$ durumuna Erler-Schnabl çözümü karşılık gelir. Daha ilginç bir çözüm \( b=-1 \) durumuna (yani  $\beta=\infty$ ) karşılık gelir.  

Daha önceki temsili, tekil gauge dönüşümleri olarak kullanarak bir çözümü diğerine eşleştirebiliriz. 
$\psi_b= U_b^{-1} QU_b$ ve $b= \frac {\beta}{1-\beta}$ olarak belirleyelim. Bu durumda
\begin{equation} 
\psi_b=  U_b^{-1} QU_b= X^{-1}\left(Q+\psi_{b'}\right) X\quad\longrightarrow\quad X= U_{b'}^{-1}U_b 
\end{equation}  
elde edilir. $b'=0$ ve $b$'yi genel alırsak 
\begin{equation} 
X= \left(1+cB \frac 1K\right)\left(1-cB \frac h{K+1}\right)= 1- cB \,\beta, \quad\quad X^{-1}= 1+cB \frac {\beta}{1-\beta} 
\end{equation} 
buluruz. Dönüşüm regular olduğu için $\beta=\infty$ ( $b=-1$) durumu hariç tüm çözümler eşdeğerdir.  

Şimdi Sen'in ikinci varsayımını kontrol edelim. Yukarıdaki $b$ (veya $\beta$) ile belirlenen çözümler için homotopi operatörünün bu şekilde olduğu kolayca gösterilebilir  
\begin{equation}  
{\cal A}= \frac 1{1+b} \,\frac B{K+1}\label{homotopy}  
\end{equation}  
Fark edilebileceği gibi, bu homotopi operatörü \( b=-1 \) için mevcut değildir. Teklif edilen çözüm Sen'in ikinci varsayımını sağlamamaktadır. Bu sebepten zamanında bu çözümü daha incelemedik, sonradan Erler bana bu çözümün fiziksel olmadığını yazdı. 

\medskip

\textbf{Teşekkürler.} Bu kısa yazı, 18-19 Nisan 2024 tarihleri arasında “Istanbul Integrability and Stringy Topics Initiative (\href{https://istringy.org/}{istringy.org})” tarafından düzenlenen Feza Gürsey Fizik Günlerinde (Boğaziçi Üniversitesi, İstanbul) yaptığım konuşmaya dayanmaktadır. Tüm organizasyon ekibine ve görüş alışverişinde bulunduğum birçok katılımcıya teşekkür etmek isterim. Makale üzerinde yaptığı iyileştirmeler için Ali Mert Yetkin'e ayrıca minnettarlığımı sunarım.

\bibliographystyle{utphys}
\bibliography{osft}

%

\end{document}